\documentclass[twoside,a4paper,10pt]{article}

\usepackage[english]{babel}
\usepackage{amsmath}
\usepackage{amssymb}
\usepackage{color}
\usepackage{amsthm}
\newcommand{\be}{\begin{equation}}
\newcommand{\ee}{\end{equation}}
\newcommand{\n}{\noindent}
\newcommand{\sgn}{\textrm{sgn}}
\newcommand{\RE}{\mathbb R}
\newcommand{\CO}{\mathbb C}
\newcommand{\SO}{\mathbb S}
\newcommand{\XX}{\mathcal X}
\newcommand{\UU}{\mathcal U}
\newcommand{\bbS}{\textrm{{\bf S}}}
\newcommand{\HH}{\mathcal H}
\newcommand{\KK}{\mathcal K}

\newcommand{\usig}{{\underline\sigma}}

\newcommand{\ualp}{{\underline\alpha}}

\renewcommand{\Im}{\textrm{Im}\,}

\newtheorem{theorem}{Theorem}
\theoremstyle{remark}

\newtheorem{example}{Example}

\title{Spin dependent point potentials in one and three dimensions} 
\author{Claudio Cacciapuoti$^{1}$, Raffaele Carlone$^{2}$,
  Rodolfo Figari$^{2}$} 
\date{}

\begin{document}
\maketitle
\begin{center}
$^1$Institut f\"ur Angewandte Mathematik der Universit\"at Bonn.\\
Wegelerstrasse 6, D-53115 Bonn, Germany.\\
\vspace{0,5cm}
$^2$Istituto Nazionale di Fisica Nucleare (INFN), Sezione di Napoli.\\
Dipartimento di Scienze Fisiche, Universit\`a di Napoli Federico II.\\
Via Cintia 80126 Napoli, Italy.\\
\end{center}
\n E-mail: caccia@na.infn.it, carlone@na.infn.it, figari@na.infn.it.
\vspace{1,5cm}
\begin{abstract}
\n We consider a system realized with one spinless quantum particle
and an array of $N$ spins 1/2 in dimension one and three. We
characterize all the Hamiltonians obtained as point 
perturbations of an assigned free dynamics
in terms of some generalized boundary conditions. For every
boundary condition we give the explicit formula for the resolvent of
the corresponding Hamiltonian. We discuss the
problem of locality and give two examples of spin dependent point
potentials that could be of interest as multi-component solvable
models.
\end{abstract}

\section{Introduction}
\noindent
Point interactions were introduced in the early days of Quantum
Mechanics in order to describe the low energy dynamics of a quantum 
particle subject to short-range forces, see, e.g., \cite{BP},
\cite{F}, \cite{KP}
and  \cite{T}. The appearance of divergent  
terms in a formal perturbation scheme using delta-like potentials was
often bypassed considering only the first term in the
expansion. 
Methods and results of the application of this kind of potentials 
to the theory of neutron scattering by solids and fluids can be found
in \cite{L}.  

\noindent
The work of Berezin and Faddeev \cite{BF} at the beginning of the
sixties opened  
the way to a complete characterization of point interaction
Hamiltonians in any dimension (for an exhaustive review of what is 
currently known about these kind of solvable models see, e.g.,
\cite{AGH-KHII}). Few years later Minlos and Faddeev \cite{MF} were
the first to point out the 
difficulties to extend zero range interactions to systems of more 
than two particles. As an aside we want to mention that neither a 
definite way-out of this ultraviolet problem in non-relativistic 
Quantum Mechanics nor a no-go theorem has been found yet. For this 
reason the range of applicability of point interactions remained 
limited to the framework of one-particle Quantum Mechanics.

\noindent
Nowadays there is a growing interest in multi-component quantum
systems and in particular in the study of the dynamics of a
microscopic
quantum system in interaction with a quantum environment. 
The evolution of the entanglement system-environment and the onset 
of the transition to a more classical behavior of the microscopic
system 
as a consequence of the interaction with the environment are the
dynamical features under analysis.

\noindent
In the following, making use of recent techniques in the theory of 
self-adjoint extensions of symmetric operators, we construct models 
for the dynamics of one quantum particle in interaction with any number 
of localized spins. In this way we are able to define simple, but
genuinely multi-component, quantum systems where conjectures and 
qualitative results in the theory of quantum open systems can, 
in principle, be rigorously approached. 

\noindent
For the sake of simplicity we examine systems consisting of one
spinless particle in interaction with localized 1/2 spins (in units
where $\hbar = 1$).  Physical 
phenomenology would suggest considering the particle with spin and a 
spin-spin interaction conserving the total spin. It is easy 
to convince oneself that, in the latter case, inside each channel
characterized by a fixed value of the total spin, the dynamics would 
be described by some Hamiltonian of the type we consider here,
possibly relative to a value of the spin larger than 1/2. Few examples of such Hamiltonians were already heuristically found and
used to study different problems, e.g., the  spin dependent
scattering \cite{L}
or the  interaction of one quantum particle with one or (several)
quantum dots \cite{ADKS} (see also \cite{SEPVS} for one example in two dimensions). The
straightforward generalization to higher values of the spin will 
not be given here. 

\noindent
In Section \ref{section2} we introduce some notation and define the
free quantum  
dynamics for the particle and the spins. In Section \ref{section3} we state and 
prove our main results: we give a complete characterization of all 
zero-range perturbations of the free dynamics in dimension one and 
three. At the end of Section \ref{section3} we discuss  with more
detail  two  examples of spin-dependent point interactions that, in our
opinion, are of interest as non trivial solvable
models. In order to make clearer our formulas, the resolvent in the
simple case of $N=1$ and $d=3$ is written in an extended form. A
section of conclusions follows.   

\section{Some notation and the free dynamics\label{section2}}

\n In this section we define the state space for a quantum system consisting
of one particle and an array of $N$ spins. Moreover we introduce some
notation and define the non-interacting Hamiltonian $H$.

\n We will consider here the case of spin 1/2. The state of
each spin placed in a fixed position of space is represented by a
unitary vector in
$\CO^2$.

\n Consider the first Pauli matrix, $\hat\sigma^{(1)}_j$, where the
index $j=1,\dots,N$ indicates that such operator refers to the
$j$-th spin. We indicate with $\chi_{\sigma_j}$ the normalized
eigenvector of the operator $\hat\sigma^{(1)}_j$ with
eigenvalue $\sigma_j=\pm1$
\be\hat\sigma^{(1)}_j\chi_{\sigma_j}=\sigma_j\chi_{\sigma_j}\qquad
\sigma_j=\pm1\,;\;\|\chi_{\sigma_j}\|_{\CO^2}=1\,;\;j=1,\dots,N\,.
\ee \n With this notation the state of the $j$-th spin can be written
as the linear superposition $a_j\,\chi_++b_j\,\chi_-$, with
$a_j,b_j\in\CO$ and $|a_j|^2+|b_j|^2=1$.

\n The natural Hilbert space for the description of a system of one
particle in dimension $d$ and $N$ spins 1/2 is then

\be \HH=L^2(\RE^d)\otimes\SO_N\,, \ee 
\n where
\be
\SO_N=\overset{N}{\overbrace{\CO^2\otimes\dots\otimes\CO^2}}
\ee

\n In this paper we will consider only the cases $d=1,3$. We indicate
with a capital Greek letter a generic vector in $\HH$.

\n Let us define $\XX_{\usig}=\chi_{\sigma_1}\otimes\dots
\otimes\chi_{\sigma_{N}}$, where $\usig$ is the N-dimensional vector
$\usig=(\sigma_1,\dots,\sigma_N)$. Trivially $\XX_\usig\in\SO_N$,
$\|\XX_\usig\|_{\SO_N}=1$ and the following decomposition formula
holds 
\be\label{decomposition}
\Psi=\sum_{\usig}\psi_\usig\otimes\XX_\usig\qquad\Psi\in\HH\,, 
\ee 
\n where the sum
runs over all the possible configurations of the vector $\usig$
while $\psi_\usig\in L^2(\RE^d)$ $\forall\usig$ is referred to as the
wave function component of the state $\Psi$. The
choice of the $\XX_\usig$ as basis of $\SO_N$ is arbitrary, we
consider the basis 
of eigenvectors of $\hat\sigma^{(1)}_j$ according to what will be our choice for
the free Hamiltonian.

\n The scalar product in $\HH$ is defined in a natural way by \be
\langle\Psi,\Phi\rangle=\sum_{\usig}(\psi_{\usig},\phi_{\usig})_{L^2}
\qquad\Psi,\Phi\in\HH\,.
\ee

\n Consider the operator in $\SO_N$ 
\be
\bbS_j=\overset{N}{\overbrace{\mathbb{I}_{\CO^2}\otimes\dots
\otimes{\hat\sigma^{(1)}_j}\otimes\dots\otimes\mathbb{I}_{\CO^2}}}
\qquad j=1,\dots,N\,.
\ee 
\n Vectors $\XX_\usig$ are eigenvectors of
$\bbS_j$, 
\be \bbS_j\XX_{\usig}=\sigma_j\XX_{\usig}\qquad
j=1,\dots,N\,.
\ee

\n The  following operator is self-adjoint in $\HH$
\be\label{DH} D(H)=H^2(\RE^d)\otimes\SO_N 
\ee 
\be\label{H}
H=-\frac{\hbar^2}{2m}\Delta\otimes\mathbb{I}_{\SO_N}+\sum_{j=1}^N\mathbb{I}_{L^2}\otimes\alpha_j\bbS_j\qquad\alpha_j\in\RE\,,
\ee 
\n here $H^2(\RE^d)$ indicates the standard Sobolev space of functions
in $L^2(\RE^d)$, with first and second
generalized derivative in $L^2(\RE^d)$. $m$ indicates the mass of the
particle and $\alpha_j$ are real 
constants with the dimension of an energy. The operator $H$ defines the
free Hamiltonian. In the following we will fix $\hbar=1$ and $2m=1$.

\n By using the decomposition formula (\ref{decomposition}) it is
easily seen that the action of $H$ on vectors in its domain is given
by 
\be 
H\Psi=\sum_{\usig}\big(-\Delta+\ualp\,\usig\big)
\psi_\usig\otimes\XX_\usig 
\qquad\Psi\in\HH\,,
\ee 
\n where $\ualp$ is the
N-dimensional real vector $(\alpha_1,\dots,\alpha_N)$ and
$\ualp\,\usig=\sum_{j=1}^N\alpha_j\sigma_j$.

\n The resolvent of $H$, $R(z)=(H-z)^{-1}$, is 
\be
R(z)\Psi=\sum_{\usig}\big(-\Delta-z+\ualp\,\usig\big)^{-1}
\psi_\usig\otimes\XX_\usig\qquad\Psi\in\HH;\;
z\in\rho(H)\,,\ee
\n where $\rho(H)$ indicates the resolvent set of $H$. We indicate
with $G^w(x-x')$ the integral kernel of the operator
$\big(-\Delta-w\big)^{-1}$. Its explicit expression is well known and reads 
\be\label{Gz}
G^w(x)=\left\{\begin{aligned}
&i\frac{e^{i\sqrt{w}|x|}}{2\sqrt{w}}
\qquad& d=1\\ \\
&\frac{e^{i\sqrt{w}|x|}}{4\pi|x|}& d=3
\end{aligned}\right.\quad w\in\CO\backslash\RE^+;\;\Im(\sqrt{w})>0
\ee
\n From the
spectral properties of the operator $-\Delta$,
with domain $D(-\Delta)=H^2(\RE^d)$, it is easily seen that the
spectrum of $H$ is 
only absolutely 
continuous, in particular
\be
\sigma_{pp}(H)=\varnothing\;;\quad\sigma_{ess}(H)=\sigma_{ac}(H)=[\mu,\infty),\;
\quad\mu=\min_{\usig}(\ualp\,\usig)\,. \ee

\n The solution of the Schr\"odinger equation 
\be
i\frac{d}{dt}\Psi^t=H\Psi^t\,,
\ee
\n with initial datum
\be
\Psi^{t=0}=\Psi^{0}=\sum_\usig\psi_\usig^0\otimes\XX_\usig
\quad\Psi^0\in\HH\,,
\ee
\n is formally written as $e^{-itH}\Psi_0$. By using the property of
the Laplace transform  
$\mathcal L^{-1}\Big(\mathcal L(f)(\cdot+s)\Big)(\tau )=e^{−s\tau}f (\tau)$ 
we obtain the strongly continuous unitary
group $e^{-itH}$ (see, e.g., Th. VIII.7 \cite{RSI})
\be
\Psi^t=e^{-iHt}\Psi^0=\sum_\usig U^t\psi_\usig^0\otimes 
e^{-i\ualp\,\usig t}\XX_\usig\,,
\ee
\n where $U^t:L^2(\RE^d)\to L^2(\RE^d)$ is the generator of the
free dynamics for one particle in $d$ dimensions
\be
(U^tf)(x)=\frac{1}{(4\pi it)^{d/2}}
\int_{\RE^d}e^{i\frac{|x-x'|^2}{4t}}f(x')dx'\,.
\ee 

\n The Hamiltonian $H$ does not give rise to any interaction among the
particle and the spins and of the spins among 
themselves.

\section{Point perturbations of $H$\label{section3}}

\n In this section we use the theory of self-adjoint extensions of
symmetric operators 
to derive the whole family of Hamiltonians that 
coincide with $H$ on functions whose support does not contain  
the set of points where the spins
are placed (for an introduction to the standard von
Neumann's theory of self-adjoint extensions of symmetric operators see,
e.g., \cite{AGII} and \cite{RSII}).

\n Let us indicate with $Y$ the set $\{y_1,\dots,y_N\}$, where
$y_j\in\RE^d$ indicates the position of the $j$-th
spin 1/2. Consider the symmetric operator on $\HH$
\be\label{H0dom}
D(H_0)=C_0^\infty(\RE^d\backslash Y)\otimes\SO_N
\ee
\be\label{H0}
H_0=-\Delta\otimes\mathbb{I}_{\SO_N}+\sum_{j=1}^N\mathbb{I}_{L^2}\otimes\alpha_j\bbS_j
\qquad\alpha_j\in\RE
\ee
\n Let  $\KK_z(H_0)=\textrm{Ker}[H_0^*-z]$ with $\Im(z)\neq0$, where
$^*$ indicates the adjoint. 
To evaluate the deficiency indices of $H_0$,
$n_+(H_0)=\textrm{dim}[\KK_i]$ and $n_-(H_0)=\textrm{dim}[\KK_{-i}]$, we 
have to find all the independent solutions of the equation   
\be
\label{funzionidifetto}
(H_0^*-z)\Phi^z=0\qquad z\in\CO\backslash\RE;\;\Phi^z\in D(H_0^*)\,. 
\ee 
\n Define
$\Phi^z=\sum_{\usig}\phi_\usig^z\otimes\XX_{\usig}$, then
equation (\ref{funzionidifetto})  is equivalent to 
\be
\Big(\phi_\usig^z,(-\Delta-\bar z+\ualp\,\usig)\psi\Big)_{L^2}=0\qquad
\phi_\usig^z\in L^2(\RE^d);\;
\forall\,\psi\in
C_0^\infty(\RE^d\backslash Y);\;z\in\CO\backslash\RE\,.
\ee

\n The  independent  solutions  of (\ref{funzionidifetto}) in $\HH$ are
\be\label{phiz1d}
\left\{
\begin{aligned}
&\Phi^{z}_{0j\usig}=G^{z-\ualp\,\usig}(\cdot-y_j)\otimes\XX_{\usig}\\
&\Phi_{1j\usig}^{z}=(G^{z-\ualp\,\usig})'(\cdot-y_j)\otimes\XX_{\usig}
\end{aligned}\right.
\qquad z\in\CO\backslash\RE\qquad d=1
\ee

\be\label{phiz3d}
\Phi_{j\usig}^z=G^{z-\ualp\,\usig}(\cdot-y_j)\otimes\XX_{\usig}\qquad
z\in\CO\backslash\RE\qquad d=3
\ee
\n where $G^w(x)$, $w\in\CO\backslash\RE^+$, is
defined in (\ref{Gz}).

\n $(G^w)'$ indicates the first derivative of $G^w$ with respect to $x$
\be\label{Gzprimo}
(G^w)'(x)=-\sgn(x)\frac{e^{i\sqrt{w}|x|}}{2}\qquad w\in\CO\backslash\RE^+;\;
\Im(\sqrt{w})>0
\quad d=1
\ee

\n Since the index $\usig$ runs over $2^N$ distinct configurations
and $j=1,\dots,N$, for $d=1$ the deficiency indices are
$n_+=n_-=N2^{N+1}$ while for $d=3$ one has $n_+=n_-=N2^{N}$. Von
Neumann's theory ensures that  self-adjoint extensions of
$H_0$ exist and they are parametrized by the unitary applications
between $\KK_i$ and $\KK_{-i}$. Accordingly the  family of operators
which are self-adjoint 
extensions of $H_0$ is characterized by $(N2^{N+1})^2$ real
parameters for $d=1$ and by $(N2^{N})^2$ real
parameters for $d=3$.

\n Let us denote with $H^\UU$ the self-adjoint extension of $H_0$
corresponding, via the von Neumann's formula, to the unitary
application 
$\UU:K_i(H_0)\to K_{-i}(H_0)$. In general, given $\UU$, it is not easy to
obtain any information
about the resolvent of $H^\UU$ and the behavior of the wave function
component of the generic vector $\Psi\in D(H^\UU)$ in the points
$y_j$.

\n Since we want to stress the relation between a given self-adjoint
operator and the coupling between the wave function and the spin
placed in $y_j$ we characterize the self-adjoint extensions in terms of
some generalized boundary conditions satisfied by the wave
function component of the vector $\Psi$. 

\n As it was shown in \cite{GG} there is a one to one correspondence
between the self-adjoint extensions of a given symmetric operator $H_0$ 
and the self-adjoint linear relations on $\CO^{m}$, where
$m=n_+(H_0)=n_-(H_0)$. Moreover,
in \cite{AP} (see also \cite{Ppp}) it was shown, in a very general
setting, that a 
generalized Krein's formula for the resolvent  exists. Such a 
formula explicitly gives the resolvent of a self-adjoint extension of
a given symmetric operator in terms of the parameters characterizing
the boundary conditions satisfied by the vectors in its
domain. Moreover the generalized formula for the resolvent given
in \cite{AP} and \cite{Ppp} avoids the problem of finding the maximal
common part of 
two extensions.

\n In this paper we use the results of \cite{AP} and \cite{Ppp} to
obtain a complete 
characterization in terms of generalized boundary conditions of all the self-adjoint extensions of the operator
$H_0$. Moreover we explicitly
give a formula for the resolvent of each self-adjoint extension of
$H_0$.

\n Let use introduce the following notation. With $\mu$ we
indicate the multi-index $\mu=(pj\usig)$ for $d=1$ and $\mu=(j\usig)$
for $d=3$. Indices $p$, $p'$, $p''$ etc. always
assume
the values 0 and 1. Indices $j$, $j'$ and so on run over
$1,\dots,N$. With $\usig$, $\usig'$, etc., we indicate $N$-dimensional
vectors, e.g.,  $(\sigma_1,\dots,\sigma_N)$
where $\sigma_j=\pm1$. As an example with this notation the vectors in
$\HH$ defined by (\ref{phiz1d}) and (\ref{phiz3d}) 
are shortly referred to as $\Phi_\mu^z$. 

\n In the following
$\delta_{i,j}$ indicates the Kronecker symbol
\be
\delta_{i,j}=\left\{\begin{aligned}
                    1&\qquad i=j\\
                    0&\qquad i\neq j
                    \end{aligned}\right.
\ee
\n moreover
\be
\delta_{\usig,\usig'}=\delta_{\sigma_1,\sigma_1'}\dots
\delta_{\sigma_N,\sigma_N'}\,.
\ee

\n Given two $m\times m$ matrices $A$ and $B$, $(A|B)$ indicates the
$m\times2m$ block matrix with the first $m$ columns given by the
columns of $A$ and the second $m$'s given by the columns of $B$. 

\begin{theorem}\label{mainth1d}($d=1$)
Define the operator 
\begin{align}
D(H^{AB})=\Big\{&\Psi=\sum_{\usig}\psi_\usig\otimes\XX_\usig\in\HH\,\Big|
\;\psi_\usig\in H^2(\RE\backslash Y)\;\forall\,\usig\,;\nonumber\\
&\sum_{\mu'}
A_{\mu,\mu'}q_{\mu'}
=\sum_{\mu'}B_{\mu,\mu'}f_{\mu'}\,;
\label{boundcond1d}
\\
&q_{0j\usig}=\psi'_\usig(y_j^-)-\psi'_\usig(y_j^+)
\,,\;q_{1j\usig}=\psi_\usig(y_j^-)-\psi_\usig(y_j^+)\,,\label{qmu1d}\\
&f_{pj\usig}
=(-)^{p}
\frac{\psi_{\usig}^{(p)}(y_{j}^+)+\psi_{\usig}^{(p)}(y_{j}^-)}{2}\,,\label{psireg1d}\\
&AB^*=BA^*\;,(A|B)\;\textrm{of maximal rank }N2^{N+1}\label{condAB1d}
\Big\}
\end{align}
\be\label{actionHTheta1d}
H^{AB}\Psi=\sum_{\usig}
(-\Delta+\ualp\,\usig)\psi_\usig\otimes\XX_\usig
\qquad\alpha_j\in\RE\,,\;x\in\RE\backslash Y\,.
\ee
\n $H^{AB}$ is self-adjoint and its resolvent,
$R^{AB}(z)=(H^{AB}-z)^{-1}$, is given by 
\be
\label{iris}
R^{AB}(z)=R(z)+\sum_{\mu,\mu',\mu''}
\big(({\Gamma^{AB}}(z))^{-1}\big)_{\mu,\mu'}B_{\mu',\mu''}
\langle\Phi^{\bar z}_{\mu''}
,\cdot\,\rangle\Phi^z_{\mu}\quad z\in\rho(H^{AB})\,.
\ee

\n Where $\Gamma^{AB}(z)$ is the $N2^{N+1}\times N2^{N+1}$ matrix 
defined as 
\be 
\Gamma^{AB}(z)=B\Gamma(z)+A\,.
\ee
with
\be\label{Gamma01d}
\begin{aligned}
&(\Gamma(z))_{pj\usig,p'j'\usig'}=0\qquad&\usig\neq\usig'\\
&(\Gamma(z))_{pj\usig,p'j\usig}=0 &p\neq p'\\
&(\Gamma(z))_{0j\usig,0j'\usig}=-G^{z-\ualp\,\usig}(y_j-y_{j'})&\\
&(\Gamma(z))_{1j\usig,1j'\usig}=-(z-\ualp\,\usig)G^{z-\ualp\,\usig}(y_j-y_{j'})&\\
&(\Gamma(z))_{1j\usig,0j'\usig}=(G^{z-\ualp\,\usig})'(y_j-y_{j'})&j\neq j'\\
&(\Gamma(z))_{0j\usig,1j'\usig}=-(G^{z-\ualp\,\usig})'(y_j-y_{j'})&j\neq j'\,.
\end{aligned}
\ee
\n Functions $G^{w}(x)$ and
$(G^{w})'(x)$ are defined in (\ref{Gz}) and
(\ref{Gzprimo}).
\end{theorem}
\begin{proof}
\n Define two linear
applications $\Lambda:D(H_0^*)\to\CO^m$ and
$\tilde\Lambda:D(H_0^*)\to\CO^m$, with $m=N2^{N+1}$. $\Lambda$ defines the charges
 $q_\mu$ in (\ref{qmu1d}) by
 \be
q_\mu=(\Lambda\Psi)_\mu\qquad\mu=(pj\usig)\;;\quad
\Psi=\sum_{\usig}\psi_\usig\otimes\XX_{\usig}\in D(H_0^*)\,.
\label{defgamma}\ee
\n $\tilde\Lambda$  defines $f_{\mu}$ in (\ref{psireg1d})
\be
f_{\mu}=(\tilde\Lambda\Psi)_{\mu}
\qquad\mu=(pj\usig)
\;;\quad
\Psi=\sum_{\usig}\psi_\usig\otimes\XX_{\usig}\in D(H_0^*)\,.
\label{defgammatilde}\ee
\n The linear functionals $\Lambda$ and $\tilde\Lambda$ correspond to
$\Gamma_1$ and $\Gamma_2$ defined in \cite{AP}. Integrating by parts
it follows that 
\be\label{relation}
\langle\Psi_1,H_0^*\Psi_2\rangle-\langle H_0^*\Psi_1,\Psi_2\rangle
=\sum_{\mu}\Big[\overline{(\Lambda\Psi_1)}_\mu(\tilde\Lambda\Psi_2)_\mu-
\overline{(\tilde\Lambda\Psi_1)}_\mu(\Lambda\Psi_2)_\mu\Big]
\ee
\n for all $\Psi_1,\,\Psi_2\in D(H_0^*)$. Moreover $\Lambda$ and
$\tilde\Lambda$ are surjective, this implies that the triple
$(\CO^m,\Lambda,\tilde\Lambda)$ is a  
boundary value space for $H_0$, see, e.g., \cite{GG}. Then from
Theorem 3.1.6 in
\cite{GG} we obtain that all the 
self-adjoint extensions 
of $H_0$  correspond to the restrictions of $H_0^*$ on vectors $\Psi$
satisfying
\be\label{salr}
\sum_{\mu'}A_{\mu,\mu'}(\Lambda\Psi)_{\mu'}=
\sum_{\mu'}B_{\mu,\mu'}(\tilde\Lambda\Psi)_{\mu'}\,,
\ee
\n where $A_{\mu,\mu'}$ and $B_{\mu,\mu'}$ are two $N2^{N+1}$ matrices
satisfying $AB^*=BA^*$ ($AB^*$ Hermitian) and $(A|B)$ with maximal rank
$N2^{N+1}$. This proves that the operators $H^{AB}$ are self-adjoint.

\n We use the proposition proved in \cite{AP} (see also Theorem 10 in
\cite{Ppp}) to  write down the resolvent of $H^{AB}$.

\n Define $\gamma_z:\CO^m\to\KK_z$ in the following way:
$\gamma_z=(\Lambda|\KK_z)^{-1}$. The action of $\gamma_z$ on a  
vector $\underline a\in\CO^m$  is given by 
\be
\gamma_z\underline a=\sum_\mu a_\mu\Phi^z_\mu
\label{gamma}
\ee
where $\Phi^z_\mu$ is defined in (\ref{phiz1d}). In fact
\be
(\Lambda \Phi^z_{p'j'\usig'})_{pj\usig}=
\delta_{\usig,\usig'}\delta_{j,j'}\delta_{p,p'}
\big[(G^{z-\ualp\,\usig})'(0^-)-(G^{z-\ualp\,\usig})'(0^+)\big]=
\delta_{\usig,\usig'}\delta_{j,j'}\delta_{p,p'}\,.
\ee

\n The adjoint of $\gamma_z$, $\gamma_z^*:\HH\to\CO^m$
is defined 
by 
\be
(\gamma_z^*\Psi)_\mu=\langle\Phi^z_\mu,\Psi\rangle
\label{gammastar}\ee
in fact
\be
\langle\Psi,\gamma_z\underline a\rangle=
\sum_{pj\usig}a_{pj\usig}
\Big(\psi_\usig(\cdot),(G^{z-\ualp\,\usig})^{(p)}(\cdot-y_j)\Big)_{L^2}=
\sum_{pj\usig}\overline{(\gamma_z^*\Psi)}_{pj\usig}a_{pj\usig}\,.
\ee
\n By straightforward calculations it is possible to show that the matrix
$\Gamma(z)=-\tilde\Lambda\gamma_z$ coincides with the definition given
in (\ref{Gamma01d}). From the definition of the domain of
$H^{AB}$ it follows that the free Hamiltonian $H$ is the self-adjoint
extension of $H_0$  corresponding to the choice  $A=1$
and $B=0$.  Then
$\gamma_z$ and $\Gamma(z)$ are analytic for $z\in\rho(H)$ and 
\be
(\Gamma(z))_{\mu,\mu'}-(\Gamma(w))_{\mu,\mu'}=(w-z)
\langle\Phi^{\bar
z}_{\mu},\Phi^{w}_{\mu'}\rangle\quad
z,w\in\rho(H)\,.
\ee

\n Making use of the result stated in \cite{AP}
(see also Theorem 10 in \cite{Ppp}) we obtain that for all
$z\in\rho(H)\cap\rho(H^{AB})$ the resolvent formula (\ref{iris})
holds. Since the resolvent of $H^{AB}$ is a finite rank perturbation of
the resolvent of $H$ we have $\sigma_{ess}(H^{AB})=\sigma_{ess}(H)=\sigma(H)$ (see,
e.g., \cite{AK}), and
$\rho(H)\cap\rho(H^{AB})=\rho(H^{AB})$.
\end{proof}

\n An analogous theorem  holds in the three dimensional case.
\begin{theorem}\label{mainth3d}($d=3$)
Define the operator
\begin{align}
D(H^{AB})=\Big\{&
\Psi=\sum_\usig\psi_\usig\otimes\XX_\usig\in\HH\Big|\;
\Psi
=\Psi^z+\sum_{\mu}q_{\mu}\Phi_{\mu}^z\,;\nonumber\\
&\Psi^z\in
D(H);\;z\in\rho(H^{AB})\,;\nonumber\\
&\sum_{\mu'}
A_{\mu,\mu'}q_{\mu'}
=\sum_{\mu'}B_{\mu,\mu'}f_{\mu'}\,;
\label{boundcond3d}
\\
&q_{j\usig}=\lim_{|x-y_j|\rightarrow 0}4\pi\,|x-y_j|\psi_{\usig}(x)
\label{qmu3d}\,,\\
&f_{j\usig}=\lim_{|x-y_j|\rightarrow 0}
\Big[\psi_\usig(x)-\frac{q_{j\usig}}
{4\pi\,|x-y_j|}\Big]\,,\\
&AB^*=BA^*\;,(A|B)\;\textrm{of maximal rank }N2^{N}\label{condAB3d}
\Big\}
\end{align}
\be\label{action3d}
H^{AB}\Psi=H\Psi^z+z\sum_{j,\usig}q_{j\usig}\Phi_{j\usig}^z
\qquad\Psi\in D(H^{AB})\,.
\ee
\n $H^{AB}$ is self-adjoint and its resolvent,
$R^{AB}(z)=(H^{AB}-z)^{-1}$, is given by 
\be
\label{iris3d}
R^{AB}(z)=R(z)+\sum_{\mu,\mu',\mu''}
\big((\Gamma^{AB}(z))^{-1}\big)_{\mu,\mu'}B_{\mu',\mu''}
\langle\Phi^{\bar z}_{\mu''}
,\cdot\,\rangle\Phi^z_{\mu}\quad z\in\rho(H^{AB})\,.
\ee
\n Where $\Gamma^{AB}(z)$ is the $N2^{N}\times N2^{N}$
matrix defined as
\be 
\Gamma^{AB}(z)=B\Gamma(z)+A\,.
\ee
\n with
\be\label{Gamma03d}
\begin{aligned}
&(\Gamma(z))_{j\usig,j'\usig'}=0\qquad&\usig\neq\usig'\\
&(\Gamma(z))_{j\usig,j\usig}=\frac{\sqrt{z-\ualp\,\usig}}{4\pi i}&\\
&(\Gamma(z))_{j\usig,j'\usig}=-G^{z-\ualp\,\usig}(y_j-y_{j'})&j\neq j'\,.
\end{aligned}
\ee
\n Function $G^{w}(x)$ is defined in
(\ref{Gz}).
\end{theorem}
\begin{proof}
\n The proof of the self-adjointness of  $H^{AB}$ is
basically the same as in the one dimensional case. Two linear, surjective
applications $\Lambda$, $\tilde\Lambda:D(H_0^*)\to\CO^m$ define the
charges $q_{j\usig}$ and the values $f_{j\usig}$ as it was
done in the one dimensional case, see (\ref{defgamma}) and
(\ref{defgammatilde}). The von Neumann decomposition formula (see, e.g.,
\cite{RSII}) gives the 
following expression for the generic vector in $D(H_0^*)$
\be
\Psi=\Psi_0+\sum_{\mu}\left(
a_{\mu}\Phi^i_{\mu}+b_{\mu}\Phi^{-i}_{\mu}
\right)\qquad a_\mu,b_\mu\in\CO;\,\Psi^0\in D(H_0)\label{vonneumann}
\ee 
with $\Phi^{\pm i}_{\mu}$ as in (\ref{phiz3d}). The action of
$H_0^*$ on its domain can be written as 
\be
H_0^*\Psi=H_0\Psi_0+i\sum_{\mu}\left(
a_{\mu}\Phi^i_{\mu}-b_{\mu}\Phi^{-i}_{\mu}
\right)\qquad a_\mu,b_\mu\in\CO;\,\Psi_0\in D(H_0)\,.
\ee
\n By using the symmetry of $H_0$ it is easily proved that,
given $\Psi_1,\Psi_2\in D(H_0^*)$ such that 
\be
\Psi_k=\Psi_{k,0}+\sum_{\mu}\left(
a_{k,\mu}\Phi^i_{\mu}+b_{k,\mu}\Phi^{-i}_{\mu}
\right)
\quad a_{k,\mu},b_{k,\mu}\in\CO;\,\Psi_{k,0}\in D(H_0),\,k=1,2
\ee
the following relation holds
\be
\begin{aligned}\label{lhs}
&\langle\Psi_1,H_0^*\Psi_2\rangle-\langle H_0^*\Psi_1,\Psi_2\rangle=\\
&=2i\sum_{j,j',\usig}(\bar a_{1,j\usig}{a}_{2,j'\usig}
-\bar b_{1,j\usig}{b_{2,j'\usig}})
\big(G^{i-\ualp\,\usig}(\cdot-y_j),G^{i-\ualp\,\usig}(\cdot-y_{j'})
\big)_{L^2}\,.   
\end{aligned}
\ee
\n On the other hand, 
\be
(\Lambda\Psi_k)_\mu=q_{k,\mu}=a_{k,\mu}+b_{k,\mu}\quad k=1,2
\ee
and 
\be
\begin{aligned}
&(\tilde\Lambda\Psi_k)_{j\usig}=
f_{k,j\usig}=
i\Bigg( a_{k,j\usig}\frac{\sqrt{i-\ualp\,\usig}}{4\pi}+ 
b_{k,j\usig}\frac{\sqrt{-i-\ualp\,\usig}}{4\pi}\Bigg)+\\
&+\sum_{j'\neq j}\Big(a_{k,j'\usig}G^{i-\ualp\,\usig}(y_{j}-y_{j'})+
b_{k,j'\usig}G^{-i-\ualp\,\usig}(y_{j}-y_{j'})\Big)\quad k=1,2\,.
\end{aligned}
\ee
\n The right hand side of relation (\ref{relation}) then reads
\be\label{rhs}
\begin{aligned}
&\sum_{\mu}\Big[\overline{(\Lambda\Psi_1)}_\mu(\tilde\Lambda\Psi_2)_\mu-
\overline{(\tilde\Lambda\Psi_1)}_\mu(\Lambda\Psi_2)_\mu\Big]=\\
&=\frac{i}{4\pi}(\bar a_{1,j\usig}a_{2,j\usig}-
\bar b_{1,j\usig}b_{2,j\usig})
(\sqrt{i-\ualp\,\usig}-\sqrt{-i-\ualp\,\usig})
+\\
&+\sum_{j'\neq j}(\bar a_{1,j\usig}a_{2,j'\usig}-
\bar b_{1,j\usig}b_{2,j'\usig})
\big(
G^{i-\ualp\,\usig}(y_{j}-y_{j'})-
G^{-i-\ualp\,\usig}(y_{j}-y_{j'})\big)\,.
\end{aligned}
\ee
\n By using the resolvent identity on
$\big(G^{i-\ualp\,\usig}(\cdot-y_j),G^{i-\ualp\,\usig}(\cdot-y_{j'})\big)_{L^2}$,
for $j\neq j'$, and by direct computation of
$\|G^{i-\ualp\,\usig}\|^2_{L^2}$ it is  shown that (\ref{lhs}) and
(\ref{rhs}) coincide. Then, also for $d=3$, the triple
$(\CO^m,\Lambda,\tilde\Lambda)$ is a boundary value space and the
restriction of $H_0^*$ to vectors satisfying (\ref{boundcond3d}) is
self-adjoint, we indicate such a restriction with $\tilde
H^{AB}$. Assume that $\Psi\in\tilde H^{AB}$ and  that it is written as
in formula (\ref{vonneumann}), posing 
\be
\Psi=\Psi^z+\sum_{\mu}q_{\mu}\Phi_{\mu}^z
\ee
with
\be 
\Psi^z=\Psi_0+\sum_\mu(a_\mu\Phi^i_\mu+b_\mu\Phi^{-i}_\mu-q_\mu\Phi^z)\,,
\ee
\n and noticing that  $q_\mu=a_\mu+b_\mu$, it follows that
$\Psi^z\in D(H)$ and that the action of $\tilde H^{AB}$ on its domain is
given by (\ref{action3d}). Then $H^{AB}$ is self-adjoint.

\n Define $\gamma_z:\CO^m\to\KK_z$ as before:
$\gamma_z=(\Lambda|\KK_z)^{-1}$. Analogously to the one dimensional
case, given a 
vector $\underline a\in\CO^m$, $\gamma_z\underline a=\sum_\mu
a_\mu\Phi^z_\mu$ (see Theorem \ref{mainth1d}). Its adjoint is
$\gamma_z^*:\HH\to\CO^m$,
$(\gamma_z^*\Psi)_\mu=\langle\Phi^z_\mu,\Psi\rangle$. As in the one
dimensional case it is  possible to show that the matrix
$\Gamma(z)=-\tilde\Lambda\gamma_z$ coincides with the definition given
in (\ref{Gamma03d}). The free Hamiltonian $H$ corresponds to the
choice $A=1$ and $B=0$, and the resolvent formula (\ref{iris3d})
follows as in the one dimensional case.
\end{proof}

\n If the matrix $B$ is invertible the generalized Krein formula
is easily reduced to the standard formula with one matrix usually
denoted with $\Theta$, see \cite{P1}.

\n The generalized boundary conditions of the form (\ref{boundcond1d}) and
(\ref{boundcond3d}) include both local and non local interactions. In
our setting local means 
that the behavior of the wave function in the point $y_j$ depends only
on the state of the 
spin placed 
in the point $y_{j}$. The sub-family of local Hamiltonians $H^{AB}$,
the only ones generally considered physically admissible, is obtained by imposing some
restrictions on the matrices
$A$ and $B$, i.e.
\begin{eqnarray}
 &d=1  &  \nonumber\\
  & & A_{pj\usig,p'j'\usig'}=B_{pj\usig,p'j'\usig'}=0\qquad\forall j\neq j'  \nonumber \\
  & &A_{pj\usig,p'j\usig'}=B_{pj\usig,p'j\usig'}=0\qquad
\textrm{if for some }\;k\neq j,\; 
\sigma_k\neq\sigma_k'\\
& &A_{pj\usig,p'j\usig'}=a_{pj\sigma_j,p'j\sigma_j'}\,;
\quad B_{pj\usig,p'j\usig'}=b_{pj\sigma_j,p'j\sigma_j'}
\quad\textrm{otherwise}\nonumber\\
&d=3&\nonumber\\
& &A_{j\usig,j'\usig'}=B_{j\usig,j'\usig'}=0\qquad\forall j\neq j'\nonumber\\
& &A_{j\usig,j\usig'}=B_{j\usig,j\usig'}=0\qquad
\textrm{if for some }\;k\neq j,\; 
\sigma_k\neq\sigma_k'\\
& &A_{j\usig,j\usig'}=a_{j\sigma_j,j\sigma_j'}\,;
\quad B_{j\usig,j\usig'}=b_{j\sigma_j,j\sigma_j'}
\quad\textrm{otherwise}\nonumber
 \end{eqnarray}
where the (complex) constants $a_{pj\sigma_j,p'j\sigma_j'}$,
$b_{pj\sigma_j,p'j\sigma_j'}$ (and $a_{j\sigma_j,j\sigma_j'}$, 
 $b_{j\sigma_j,j\sigma_j'}$) are subjected to the 
restriction (\ref{condAB1d}) (and (\ref{condAB3d})). 

\n We give the explicit form
of two local Hamiltonians  that we consider of special
interest.  

\begin{example}\label{ex1}$\delta$-like interactions. 

\n Consider the following choice for the matrices $A$ and $B$
\be\label{deltalike}
\begin{aligned}
&d=1&&d=3 \\
&a_{pj\sigma_j,p'j\sigma_j'}=\delta_{p,p'}\delta_{\sigma_j,\sigma_j'}&&
a_{j\sigma_j,j\sigma_j'}=\beta_{j\sigma_j}\delta_{\sigma_j,\sigma_j'}\\
&b_{0j\sigma_j,0j\sigma_j'}=-2\beta_{j\sigma_j}\delta_{\sigma_j,\sigma_j'}&&
b_{j\sigma_j,j\sigma_j'}=\delta_{\sigma_j,\sigma_j'}\\
&b_{pj\sigma_j,p'j\sigma_j'}=0\quad
\textrm{for }p\neq0\;\textrm{or
}p'\neq0\qquad&&\textrm{with }\beta_{j\sigma_j}\in\RE\\
&\textrm{with }\beta_{j\sigma_j}\in\RE&&
\end{aligned}
\ee
\n We indicate with $H^\delta$ the generic Hamiltonian in this 
sub-family of local interactions. For $d=1$, the wave function
component of
the generic state $\Psi\in D(H^\delta)$ is continuous but with 
discontinuous derivative, in particular the following boundary
conditions hold
\be
\begin{aligned}
\psi_\usig(y_j^+)=\psi_\usig(y_j^-)\equiv\psi_\usig(y_j)\;,
\qquad\psi'_\usig(y_j^+)-\psi'_\usig(y_j^-)
=\beta_{j\sigma_j}\psi_{\usig}(y_{j})\,.
\end{aligned}
\ee
\n For $d=3$ the boundary conditions simply read
\be
\beta_{j\sigma_j}q_{j\usig}=f_{j\usig}\,.
\ee

\n Following a practice common in the literature (see \cite{AGH-KHII}
and references  
therein), we refer to $H^\delta$ as $\delta$-like
interactions. We would like to stress that such  boundary conditions
are diagonal in the spin variables. This  means that the $\chi_+$
component of the $j$-th
spin  affects
only the wave function component relative 
to the configuration of the spins  with the $j$-th one in the state
$\chi_+$. This implies that,  
given the initial state
$\Psi^{t=0}=\psi^0\otimes\XX_\usig$, the evolution generated
by $H^\delta$ gives $\Psi^t=\psi^t\otimes\XX_\usig$. Here
$\psi^t(x)=(U_\beta^t\psi^0)(x)$, where $U_\beta^t$ is a strongly continuous
unitary group in $L^2(\RE^d)$. 

\n An analogous remark holds for all the
boundary conditions that are diagonal in the spin variables. While in
dimension three they are only of the form given in the
example, in dimension one the family of self-adjoint boundary
conditions is richer. Among
them we recall the ones corresponding to a $\delta'$ coupling (see
\cite{AGH-KHII}), whose domain consists of discontinuous wave functions
with continuous 
derivative such that the jump of the wave function in $y_j$ is 
proportional to the value of the first derivative in $y_j$.
\end{example}

\begin{example}\label{ex2}Off diagonal interactions.

\n Let us consider the local interactions defined by
\be
\begin{aligned}
&d=1&&d=3 \\
&a_{pj\sigma_j,p'j\sigma_j'}=\delta_{p,p'}\delta_{\sigma_j,\sigma_j'}&&
a_{j\sigma_j,j\sigma_j'}=\sigma_j i\hat\beta_{j\sigma_j}(1-\delta_{\sigma_j,\sigma_j'})\\
&b_{0j\sigma_j,0j\sigma_j'}=-2\sigma_j i\hat\beta_{j\sigma_j}(1-\delta_{\sigma_j,\sigma_j'})&&
b_{j\sigma_j,j\sigma_j'}=\delta_{\sigma_j,\sigma_j'}\\
&b_{pj\sigma_j,p'j\sigma_j'}=0\quad
\textrm{for }p\neq0\;\textrm{or
}p'\neq0\qquad&&\textrm{with }\hat\beta_{j\sigma_j}\in\RE\\
&\textrm{with }\hat\beta_{j\sigma_j}\in\RE&&
\end{aligned}
\ee
\n A simple calculation gives the corresponding boundary
conditions. For $d=1$
\be
\begin{aligned}
&\psi_\usig(y_j^+)=\psi_\usig(y_j^-)\equiv\psi_\usig(y_j)\\
&\psi'_\usig(y_j^+)-\psi'_\usig(y_j^-)
=\sigma_j i\beta_{j\sigma_j}\psi_{(\sigma_1\dots\sigma_j'\dots\sigma_N)}(y_{j})\quad
\sigma_j'\neq\sigma_j\,.
\end{aligned}
\ee
\n and for $d=3$
\be
\sigma_j
i\hat\beta_{j\sigma_j}q_{j(\sigma_1\dots\sigma_j'\dots\sigma_N)}=
f_{j\usig}\quad\sigma_j'\neq\sigma_j\,.
\ee
\n The class of Hamiltonians proposed in this second example are the
simplest off diagonal ones. The interaction with the
particle induces the spins to evolve towards a superposition
state also when  the initial state is such that every spin
is in an eigenstate of $\hat\sigma_j^{(1)}$,
$\Psi^{t=0}=\psi^0\otimes\XX_\usig$. 
\end{example}

\n We regard as useful to give, at least in the simplest case of one
spin, the explicit expression of the resolvent of the Hamiltonians
proposed in examples
\ref{ex1} and \ref{ex2}. This is done in the following:
\begin{example}One spin in dimension three.

\n Let us consider the case 
of one spin in dimension three placed in the point $y\in\RE^3$. We indicate
with $R^\delta(z)$ the resolvent of the Hamiltonian $H^\delta$ defined in
example \ref{ex1} when $N=1$. The resolvent  $R^\delta(z)$  can be written as
\be
\begin{aligned}
R^{\delta}(z)=&\Big[G^{z-\alpha}+
\frac{4\pi i}{\sqrt{z-\alpha}+4\pi i\beta_+}
G^{z-\alpha}(\cdot-y)G^{z-\alpha}(y-\cdot)\Big]\otimes(\chi_+,\cdot)_{\CO^2}\chi_++\\
+&\Big[G^{z+\alpha}+
\frac{4\pi i}{\sqrt{z+\alpha}+4\pi i\beta_-}
G^{z+\alpha}(\cdot-y)G^{z+\alpha}(y-\cdot)\Big]\otimes(\chi_-,\cdot)_{\CO^2}\chi_-\,.
\end{aligned}
\ee
\n The expressions in the square brackets are identical to the
resolvent of the operator formally written as ``$-\Delta+\beta_\sigma\delta_y$''
in dimension three (see \cite{AGH-KHII}). Then  all the results
concerning the delta-potential in dimension three can be adapted to
$H^\delta$. Let us recall that the generator of the dynamics can be
formally written as $e^{-iH^\delta t}=-\mathcal
L^{-1}\big((H^\delta-\cdot)^{-1}\big)(-it)$, then,
due to the presence of the
projectors $(\chi_+,\cdot)_{\CO^2}\chi_+$ and $(\chi_-,\cdot)_{\CO^2}\chi_-$ the
dynamics generated by $H^\delta$ is factorized in the spin components. 

\n Let us indicate with $H^{od}$ the Hamiltonian corresponding to the
one defined in
example \ref{ex2}, in dimension three and with $N=1$. Its resolvent can be
explicitly written with the following large formula 
\be
\begin{aligned}
&R^{od}(z)=G^{z-\alpha}\otimes(\chi_+,\cdot)_{\CO^2}\chi_++
G^{z+\alpha}\otimes(\chi_-,\cdot)_{\CO^2}\chi_-+\\
&-\frac{4\pi i\sqrt{z+\alpha}}
{(4\pi)^2\hat\beta_+\hat\beta_--\sqrt{z-\alpha}
\sqrt{z+\alpha}}G^{z-\alpha}(\cdot-y)G^{z-\alpha}(y-\cdot)
\otimes(\chi_+,\cdot)_{\CO^2}\chi_++\\
&-\frac{4\pi i\sqrt{z-\alpha}}
{(4\pi)^2\hat\beta_+\hat\beta_--\sqrt{z-\alpha}
\sqrt{z+\alpha}}G^{z+\alpha}(\cdot-y)G^{z+\alpha}(y-\cdot)
\otimes(\chi_-,\cdot)_{\CO^2}\chi_-+\\
&-\frac{i\hat\beta_+}
{(4\pi)^2\hat\beta_+\hat\beta_--\sqrt{z-\alpha}
\sqrt{z+\alpha}}G^{z-\alpha}(\cdot-y)G^{z+\alpha}(y-\cdot)
\otimes(\chi_-,\cdot)_{\CO^2}\chi_++\\
&+\frac{i\hat\beta_-}
{(4\pi)^2\hat\beta_+\hat\beta_--\sqrt{z-\alpha}
\sqrt{z+\alpha}}G^{z+\alpha}(\cdot-y)G^{z-\alpha}(y-\cdot)
\otimes(\chi_+,\cdot)_{\CO^2}\chi_-+
\end{aligned}
\ee
\n The terms $(\chi_-,\cdot)_{\CO^2}\chi_+$ and $(\chi_+,\cdot)_{\CO^2}\chi_-$
indicate that, in such a case, the dynamics cannot be factorized in the spin
components. Furthermore there are not ``ready to use'' formulas that
can be used  to evaluate the  spectrum or  the
propagator of $H^{od}$.  
\end{example}

\section{Conclusions}

\n In the previous sections we introduced a family of Hamiltonians
describing the dynamics of a quantum system consisting of one particle 
in interaction with an array of localized spins.

\noindent 
Different self-adjoint extensions of the free Hamiltonian correspond to 
different physical models of interaction between the particle and the
spins. 
In fact it is possible to characterize particular subfamilies of
extensions according to different features of the dynamics they generate. 

\noindent
In example \ref{ex1}  we  identified the  
sub-family of $\delta$-like Hamiltonians. While the spin dynamics is
unaffected by the 
interaction, the particle ``feels'' zero-range forces whose strength
depends on the value of some spin component of the localized spin. 
Those interaction models are a rigorous version of the spin-dependent 
delta potentials that have been one of the main tool in the
description of neutron scattering by condensed matter \cite{L}. 

\noindent
Our current aim is to build up simple models for a quantum measurement 
apparatus detecting ``the trajectory'' of a quantum particle. Mott first
considered this problem in a seminal paper \cite{Mo}. He was looking for an 
explanation of the appearance of sharp classical-like tracks in
particle detectors in high energy Physics experiments. Mott's paper
remained 
almost unnoticed till the second half of the last century when a 
renewed interest in the measurement problem showed up in the community 
of theoretical physicists. Since that time the possibility to
understand at least some qualitative features of the measurement
process thoroughly inside the framework of Quantum Mechanics, 
without rely on any ``reduction of the wave packet'' postulate, has been 
matter of debate in fundamental and applied Theoretical Physics (see,
e.g., \cite{HS}, \cite{JZ}, \cite{DFT}, \cite{AFFT}, \cite{AFFT05}, \cite{CCF}).

\noindent
The first attempt to analyze, in a simple setting, the dynamics of 
a quantum particle interacting with a many body quantum system is due 
to Hepp  (\cite{H} see also \cite{Se} for recent results on the subject) . 
He defined a one dimensional model (often referred to as the 
Coleman-Hepp model) of a quantum measurement apparatus suitable 
for the measure of the spin of a particle through its interaction with 
an array of localized spins. In order to simplify the treatment the
particle wave function was supposed to translate with constant
velocity according to a free non-dispersive dynamics. It is worth 
mentioning that the Hamiltonians described in example \ref{ex1} 
might be used to define a completely quantum Coleman-Hepp model. 

\noindent
Following the original idea of Mott we started to analyze models
similar to the one described by Hepp, where the dynamics of the spins 
is significantly affected by the particle wave function. The
Hamiltonians described in example \ref{ex2} makes available a solvable
model where rigorous results on the 
dynamics of a quantum particle in (a simplified version of) a  particle 
detector might be obtained.\\

{\bf Acknowledgments:} The authors thank Andrea Posilicano for
useful exchanges of views about the theory of self-adjoint extensions
of symmetric operators. We thank also Alessandro Teta and Gianfausto
Dell'Antonio  for all the enlightening discussions about Mott's
paper. This work was 
supported by the 
EU-Project ``Quantum Probability with Applications to Physics,
Information Theory and Biology''.
 
%BIBLIOGRAPHY
\bibliographystyle{amsplain}
\bibliography{Nspins.bib}
\end{document}